\documentclass[reprint,superscriptaddress,showpacs,amsmath,amssymb,aps,prl]{revtex4-1}

\usepackage{graphicx}
\usepackage{dcolumn}
\usepackage{bm}
\usepackage{hyperref}
\usepackage{xcolor}
\mathchardef\mhyphen="2D

\begin{document}


\title{Signatures of a Pressure-Induced Topological Quantum Phase Transition in BiTeI}

\author{Xiaoxiang Xi}
\affiliation{Photon Sciences, Brookhaven National Laboratory, Upton, New York 11973, USA}

\author{Chunli Ma}
\affiliation{Geophysical Laboratory, Carnegie Institution of Washington, Washington D.C. 20015, USA}
\affiliation{State Key Laboratory of Superhard Materials, Jilin University, Changchun 130012, China}

\author{Zhenxian Liu}
\affiliation{Geophysical Laboratory, Carnegie Institution of Washington, Washington D.C. 20015, USA}

\author{Zhiqiang Chen}
\affiliation{Department of Geosciences, Stony Brook University, Stony Brook, New York 11794, USA}

\author{Wei~Ku}
\affiliation{Condensed Matter Physics and Materials Science Department, Brookhaven National Laboratory, Upton, New York 11973, USA}

\author{H. Berger}
\affiliation{Institute of Condensed Matter Physics, \'{E}cole Polytechnique F\'{e}d\'{e}rale de Lausanne, CH-1015 Lausanne, Switzerland}

\author{C. Martin}
\affiliation{Department of Physics, University of Florida, Gainesville, Florida 32611, USA}

\author{D. B. Tanner}
\affiliation{Department of Physics, University of Florida, Gainesville, Florida 32611, USA}

\author{G. L. Carr}
\affiliation{Photon Sciences, Brookhaven National Laboratory, Upton, New York 11973, USA}
\date{\today}

\begin{abstract}
We report the observation of two signatures of a pressure-induced topological quantum phase transition in the polar semiconductor BiTeI using x-ray powder diffraction and infrared spectroscopy. The \mbox{x-ray} data confirm that BiTeI remains in its ambient-pressure structure up to 8~GPa. The lattice parameter ratio $c/a$ shows a minimum between 2.0--2.9~GPa, indicating an enhanced $c$-axis bonding through $p_z$ band crossing as expected during the transition. Over the same pressure range, the infrared spectra reveal a maximum in the optical spectral weight of the charge carriers, reflecting the closing and reopening of the semiconducting band gap. Both of these features are characteristics of a topological quantum phase transition, and are consistent with a recent theoretical proposal.
\end{abstract}
\pacs{64.70.Tg, 71.70.Ej, 78.20.-e, 61.05.cp}
\maketitle

\begin{figure*}[t]
\includegraphics[scale=0.95]{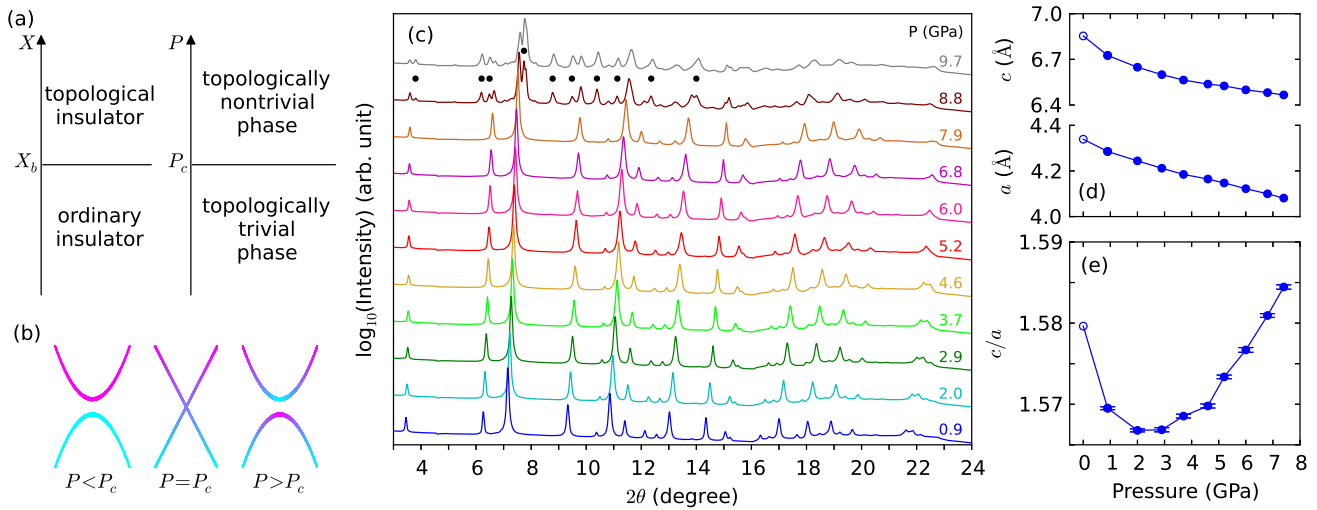}   
\caption{(color online). (a) Left: A topological insulator in contact with an ordinary insulator. $X$ is the spatial coordinate. $X_b$ indicates the physical boundary that hosts the gapless metallic state. Right: A topological quantum phase transition induced by tuning a parameter $P$ (e.g., pressure). $P_c$ indicates the critical point where the transition occurs, at which the band gap closes to yield a metallic state. (b) A diagram illustrating the band inversion across the TQPT at $P_c$. Magenta denotes the Bi-6$p_z$ orbital, cyan denotes the Te-5$p_z$ and I-5$p_z$ orbitals. (c) X-ray powder diffraction patterns of BiTeI from 0.9 to 9.7~GPa. The data above 0.9~GPa are vertically shifted for clarity. The dots indicate the new peaks that emerge above 7.9~GPa. The filled circles in (d) and (e) are the lattice parameters $c$, $a$, and the $c/a$ ratio in the hexagonal phase at different pressures, extracted from our XRD data. The open circles are the ambient-pressure data taken from Ref~\cite{Shevelkov1995}. Error bars in (d) are within the symbol size. Error bars in (e) come from the uncertainties in $c$ and $a$. } 
\label{FIG1}
\end{figure*}

Topological insulators are a class of materials having nontrivial topology in their bulk electronic states, characterized by nonzero topological invariants \cite{Hasan2010}. This intriguing state of matter was theoretically predicted \cite{Bernevig2006,Fu2007,Zhang2009,Yan2010,Lin2010} and experimentally observed \cite{Konig2007,Hsieh2008,Hsieh2009,Xia2009,Chen2009,Chen2010,Brune2011} in a variety of 2D and 3D systems. When in contact with an ordinary (i.e., topologically trivial) insulator, the band gap of the topological insulator inevitably becomes closed at the boundary where the topological invariant has a discontinuity from a nonzero value to zero, the result is a topologically protected metallic boundary state. The same principle also guarantees the band gap closing in the bulk and the resulting metallic bulk state at the critical point of a topological quantum phase transition (TQPT) [see Fig.~\ref{FIG1}(a)], hinted in BiTl(S$_{1-x}$Se$_x$)$_2$ \cite{Xu2011,Sato2011} and (Bi$_{1-x}$In$_x$)$_2$Se$_3$ \cite{Brahlek2012,Wu2013} by doping. The effects of doping in these experiments are to tune the spin-orbit interaction and/or the lattice parameters. Such effects can be realized directly by pressure, a method that does not involve the potential defects and inhomogeneity of doping. This has indeed been proposed as a means to look for topological insulating states in, e.g., ternary Heusler compounds~\cite{Chadov2010}, $A_2$Ir$_2$O$_7$ ($A=\mathrm{Y}$ or rare-earth elements)~\cite{Yang2010}, Ge$_2$Sb$_2$Te$_5$ \cite{Sa2011}, and BiTeI \cite{Bahramy2012}.

Despite its importance, there has been no experimental evidence for a pressure-induced TQPT. In general, a direct detection of the topological invariant is difficult because of its lack of coupling to most experimental probes. Consequently, a common approach is to detect the gapless metallic surface state as an indication of the topological phase using surface-sensitive techniques such as angle-resolved photoemission spectroscopy \cite{Hsieh2008,Hsieh2009,Xia2009,Chen2009,Chen2010,Brune2011} and scanning tunneling microscopy or spectroscopy \cite{TZhang2009,Alpichshev2010}. At present these techniques cannot be implemented easily at high pressure, rendering them incapable of studying pressure-induced TQPTs. Transport measurements can also detect topological surface states and are compatible with high-pressure techniques, but naturally grown materials often contain impurity-induced doping that causes conduction in the bulk to overwhelm the surface conductivity \cite{Checkelsky2009,Peng2010,Analytis2010}. Expecting the band gap closing at the critical point of TQPTs, we propose to establish pressure-induced TQPTs by identifying such band gap closing and the resulting metallic bulk state.

In this Letter, we report experimental observation of characteristics of the pressure-induced TQPT in BiTeI using x-ray powder diffraction and infrared spectroscopy. X-ray powder diffraction shows that BiTeI remains in the hexagonal structure up to 8~GPa but with the lattice parameter ratio $c/a$ passing through a minimum between 2.0--2.9 GPa, indicating an enhanced $c$-axis bonding through $p_z$ band crossing when a metallic bulk state occurs. Infrared spectroscopy over the same pressure range reveals a maximum in the free-carrier spectral weight as would be expected when the semiconducting band gap closes and reopens. Both of these results provide evidence for a pressure-induced TQPT in BiTeI at a critcal pressure $P_c$ between 2.0--2.9 GPa. Our experiments demonstrate pressure as an efficient way to induce TQPTs and the use of infrared spectroscopy and x-ray diffraction (XRD) for their investigation.  

BiTeI is a layered polar semiconductor with the trigonal space group $P3m1$ and a hexagonal unit cell at ambient pressure~\cite{Shevelkov1995}. The triple Te-Bi-I layers stack along the crystallographic $c$ axis, coupled by the van der Waals interaction. The broken inversion symmetry in the presence of strong spin-orbit coupling gives rise to a giant Rashba-type spin splitting, the largest observed so far in any system \cite{Ishizaka2011}. Although BiTeI is a narrow-gap semiconductor with an ambient-pressure band gap of $\sim$0.38~eV~\cite{Ishizaka2011,Lee2011,Martin2012}, it is typically self-doped due to nonstoichiometry and thus an $n$-type semiconductor \cite{Kanou2013}. Remarkably, Bahramy \textit{et al.} \cite{Bahramy2012} predicted that hydrostatic pressure effectively tunes the crystal-field splitting and spin-orbit interaction in BiTeI, turning it into a topological insulator. Such a TQPT is accompanied by a band inversion near the $A$ point of the Brillouin zone; the dispersion of the bulk spin-split bands becomes almost linear at the critical pressure $P_c$.


Single crystals of BiTeI were grown by the Bridgman method. The ambient-pressure carrier density is estimated to be 2.7$\times 10^{19}$~cm$^{-3}$ \cite{Martin2013}. Samples used in different experiments were mechanically cleaved from one single crystal. High-pressure infrared spectroscopy and angle-dispersive x-ray powder diffraction were performed at U2A and X17C beamlines at the National Synchrotron Light Source (Brookhaven National Laboratory). All data were collected at room temperature. Pressure was monitored by laser-excited ruby fluorescence. The separation of the two fluorescence lines indicated that the pressure was quasihydrostatic. In the XRD experiment, a sample was ground into fine powder and loaded into a diamond anvil cell. A 4:1 methanol-ethanol mixture was used as the pressure-transmitting medium. Two-dimensional diffraction rings were collected for pressures up to 27.2~GPa, with the incident monochromatic x-ray wavelength set to 0.4066~\AA. Integrating the diffraction rings yielded XRD patterns as a function of the diffraction angle $2\theta$ \cite{Hammersley1996}. 

\begin{figure}[t]
\includegraphics[scale=0.95]{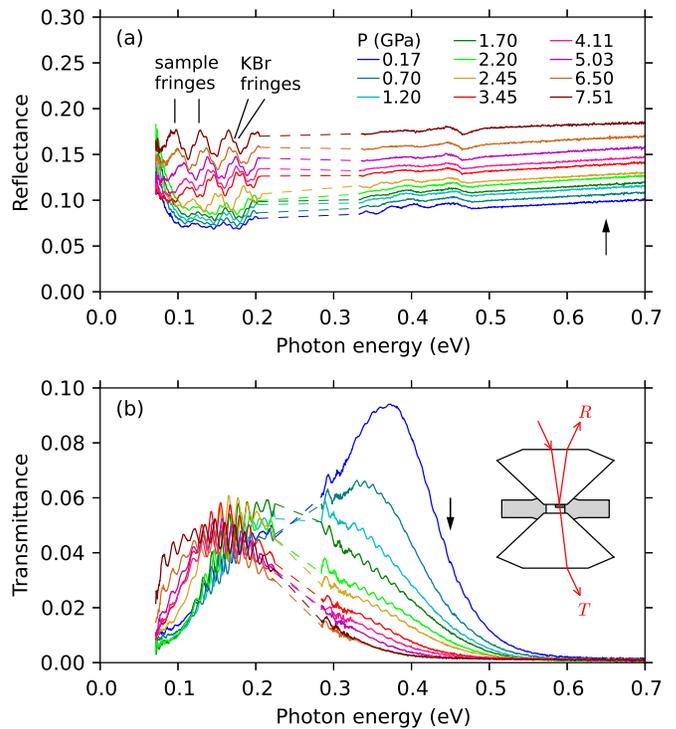}   
\caption{(color online). Infrared reflectance (a) and transmittance (b) of BiTeI for pressure in 0.17--7.51~GPa. The black arrows indicate the direction of increasing pressure. Absorption from diamond corrupts the data between 0.22--0.33~eV (patched as dashed lines) and causes the dip at $\sim$0.46~eV in reflectance. The inset in (b) shows the measurement configuration. The arrowed lines indicate the beam paths for the reflected ($R$) and transmitted ($T$) signals.} 
\label{FIG2}
\end{figure}

The experimental configuration for the infrared measurements is illustrated in the inset of Fig.~\ref{FIG2}(b). The sample was pressed against a diamond anvil and surrounded by a pressure-transmitting medium (KBr) elsewhere. Infrared microspectroscopy was performed on a Fourier transform infrared spectrometer coupled to a microscope \cite{SupMat}. Reflectance and transmittance were measured for photon energies between 0.08--1.00 eV and pressures up to 25.2~GPa. Four samples were measured, all giving consistent results. The data with the most complete pressure dependence are presented here.


The XRD patterns in Fig.~\ref{FIG1}(c) show a structural phase transition at $\sim$8~GPa. Ambient-pressure BiTeI has a hexagonal lattice with lattice parameters $a=4.339$~\AA~and $c=6.854$~\AA~\cite{Shevelkov1995}. Increasing pressure consistently shifts each Bragg peak to a larger $2\theta$ angle, indicating the reduction of the $d$ spacing. Above 7.9~GPa, a great number of new peaks emerge, suggesting a different structure. Additional measurements with smaller pressure steps confirm no structural variation below 8~GPa \cite{SupMat}. Rietveld refinement \cite{Larson2000} of the XRD data yields the lattice parameters [Fig.~\ref{FIG1}(d)] of the hexagonal phase. We include the ambient-pressure values of these parameters taken from Ref~\cite{Shevelkov1995} (omitted in our experiment), shown as open circles in Fig.~\ref{FIG1}(d); they are consistent with the pressure dependence of our data. We note in passing that a third structure appears above $\sim$18~GPa.

The ratio $c/a$ shown in Fig.~\ref{FIG1}(e) has a clear minimum between 2.0--2.9~GPa, strongly indicating a change in the chemical bonding across the TQPT. In a topologically trivial insulator, a TQPT occurs when the atomic orbital energy ordering near the Fermi level reverses and the conduction (valence) band inverts. At the critical pressure $P_c$ the band gap closes, giving rise to a bulk metallic state. The diagram in Fig.~\ref{FIG1}(b) illustrates this process. In ambient-pressure BiTeI, the bottom conduction bands are dominated by the Bi-6$p_z$ orbital while the top valence bands are dominated by the Te-5$p_z$ (and I-5$p_z$) orbital \cite{Bahramy2012,Bahramy2011}; within the low-energy bands, the bonding is almost ionic in nature for $P<P_c$. After the pressure-induced band inversion happens ($P>P_c$), one expects the Te-5$p_z$ orbital contributes to the bottom conduction bands and the Bi-6$p_z$ orbital dominates the top valence bands; the bonding becomes more covalent. At the transition ($P = P_c$), additional metallic bonding occurs involving the $p_z$ orbitals of Bi and Te, whose charge fluctuations would make the $c$ axis more compressible. This is observed clearly as a minimum in the ratio $c/a$, consistent with the notion that electronically metals are compressible but insulators are not. The minimum indicates a TQPT at $P_c$ between 2.0--2.9~GPa. The corresponding volume contraction is 7\%--9\%, similar to the theoretically predicted 11\% \cite{Bahramy2012}. Since an intermediate metallic state at the transition between two insulating states is only known to occur in a TQPT, we thus have the first piece of strong evidence of a TQPT.


Infrared spectroscopy reveals more detailed information on the electronic structure in this pressure range and indicates a TQPT as well. Both the reflectance [Fig.~\ref{FIG2}(a)] and the transmittance [Fig.~\ref{FIG2}(b)] contain fringes; they arise because of Fabry-P\'{e}rot effects in the optically flat sample and in the KBr. The thinner sample gives rise to the broad fringes and the thicker KBr gives rise to the fine ones \cite{SupMat}. When pressure is increased above 2.20~GPa, the contrast of the broad fringes in reflectance increases significantly, suggesting a sudden reduction of low-energy (below 0.2~eV) absorption right above 2.20~GPa. Between 0.08--0.20~eV, the transmittance changes slowly below 2.20~GPa but jumps to a higher level at 2.45~GPa, confirming the reduction of absorption. Such pressure-induced change of the low-energy absorption becomes more apparent in the optical conductivity (shown in Fig.~\ref{FIG4}) to be discussed below. Evidently, a transition happens at 2.20~GPa, coinciding with the pressure where the minimum $c/a$ occurs. 

\begin{figure}[t]
\includegraphics[scale=0.95]{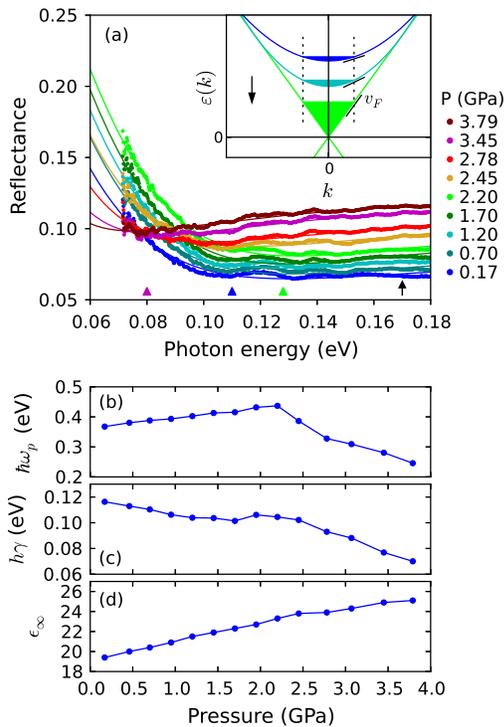}   
\caption{(color online). (a) Reflectance at low pressure and low photon energy after fringe removal. The triangles indicate the plasma edge at 0.17 GPa (blue), 2.20 GPa (lime green), and 3.45 GPa (magenta). The solid lines are fits as described in the text. The inset illustrates the conduction band dispersion $\varepsilon(k)$ and the chemical potential as the pressure approaches $P_c$. The slope of $\varepsilon(k)$ at the chemical potential gives the Fermi velocity $v_F$. The arrows indicate the direction of increasing pressure. (b)--(d) Pressure dependence of the fitting parameters $\omega_p$, $\gamma$, and $\epsilon_{\infty}$. Error bars are within the symbol size.} 
\label{FIG3}
\end{figure}

\begin{figure}[t]
\includegraphics[scale=1.1]{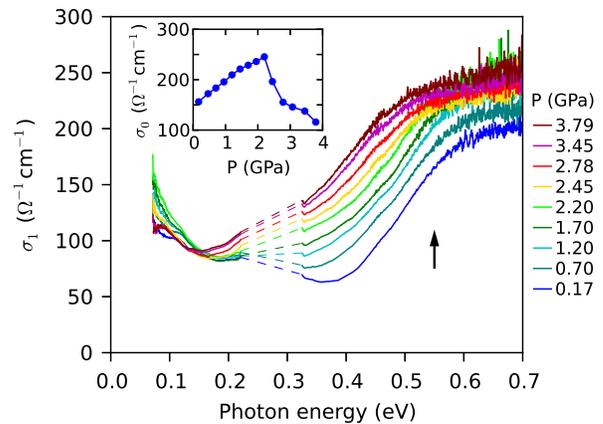}   
\caption{(color online). The real part of the optical conductivity calculated from reflectance and transmittance. The arrow indicates the direction of increasing pressure. The data between 0.22--0.33 eV are corrupted by the diamond absorption (patched as dashed lines). The inset shows the dc conductivity $\sigma_0$ calculated from $\omega_p$ and $\gamma$ shown in Figs.~\ref{FIG3}(b) and \ref{FIG3}(c).} 
\label{FIG4}
\end{figure}

We removed the fringes by Fourier transforming the spectra, filtering the fringe signatures, and taking the inverse transform. Reflectance data after fringe removal are plotted as dots in Fig.~\ref{FIG3}(a), showing a plasma edge characteristic of the free-carrier response in the system. As pressure increases from 0 to 2.20~GPa, the plasma edge gradually blueshifts. Above 2.20~GPa, it begins to redshift, with the shift continuing as the pressure is further increased. We note that the turning of the plasma edge at $\sim$2~GPa was consistently observed in the four samples we measured, although the quantitative behavior of the infrared data is sample dependent on account of differing scattering rates in the samples. Moreover, it is reversible as long as the pressure does not exceed $\sim$18~GPa, at which a second structural phase transition occurs.

The low-energy spectral feature due to the free-carrier response can be fitted with a simple Drude dielectric function
\begin{equation}
\epsilon=\epsilon_{\infty}-\frac{\omega_p^2}{\omega^2+i\omega\gamma}.\label{eqDrude}
\end{equation} 
Here $\epsilon_{\infty}$ is the dielectric constant contributed from high-energy excitations, $\omega_p^2/8$ the carrier spectral weight, and $\gamma$ the electronic scattering rate. A least-squares fit to Eq.~(S1) in~\cite{SupMat} was performed on reflectance, shown as solid lines in Fig.~\ref{FIG3}(a). The fitting parameters are shown in Fig.~\ref{FIG3}(b)--\ref{FIG3}(d) \cite{Note2}. Compared to an earlier ambient-pressure measurement \cite{Martin2012}, the scattering rate $\gamma$ is bigger in the sample studied here, and varies among all the samples measured. The monotonic increase in $\epsilon_{\infty}$ upon pressurization accords with that in reflectance at photon energies greater than 0.3~eV.


The pressure dependence of $\omega_p$ offers a second indication of a TQPT. In our $n$-type sample, the pressure-induced change in the electronic band structure affects the Fermi velocity $v_F$. As illustrated in the inset of Fig.~\ref{FIG3}(a), across a pressure-induced TQPT, $v_F$ increases as $P\rightarrow P_c$ (and decreases above $P_c$ when the dispersion again becomes parabolic, not shown). The maximum $v_F$ lies in the transition region where the band gap closes and the band dispersion approaches linear behavior. Because the carrier spectral weight $\omega_p^2/8$ simply measures $v_F^2$ over the Fermi surface in weakly interacting systems \cite{Lee2011}, one expects $\omega_p\propto v_F$, hence a maximum $\omega_p$ near $P_c$. Indeed, in Fig.~\ref{FIG3}(b) the extracted $\omega_p$ clearly peaks at 2.20~GPa, reflecting the closing and reopening of the band gap across the TQPT. The critical pressure is consistent with the range given by the x-ray powder diffraction.


The reflectance and transmittance data may be inverted to give the optical conductivity \cite{SupMat}; the results also show a maximum in the free-carrier spectral weight consistent with Fig.~\ref{FIG3}(b). The real part of the optical conductivity is shown in Fig.~\ref{FIG4}. The dc conductivity $\sigma_0$, roughly estimated by extrapolating the low-energy portion of the optical conductivity to zero photon energy, has a maximum at 2.20~GPa, consistent with the maximum in $\sigma_0 = \omega_p^2/4\pi\gamma$ calculated using the $\omega_p$ and $\gamma$ from the fit (inset in Fig.~\ref{FIG4}) and directly confirming the band gap closing and reopening during a TQPT.

In summary, we obtained independent and consistent experimental evidence for a pressure-induced topological quantum phase transition in BiTeI using x-ray powder diffraction and infrared spectroscopy. Signatures of the transition are evident in both structural data (with a minimum in the $c/a$ ratio at the critical pressure) and optical data (with a maximum in the free-carrier spectral weight at the same pressure). Both of these features are characteristics of the topological quantum phase transition and are consistent with a recent theoretical proposal. Our work motivates the exploration of the novel physics associated with this transition \cite{Yang2013} using other experimental probes.

We thank M. S. Bahramy, B.-J. Yang (RIKEN), and Jianming Bai (BNL) for useful discussions. This work was supported by the U. S. Department of Energy through Contract No. DE-AC02-98CH10886 at BNL. The use of U2A and X17C beamline was supported by NSF (Grants No. DMR-0805056 and No. EAR 06-49658, COMPRES) and DOE/NNSA (Grant No. DE-FC03-03N00144, CDAC).

\textit{Note added.}---High pressure infrared and Raman spectroscopy on BiTeI were reported in a recent preprint \cite{Tran2013}.

\clearpage

\setcounter{figure}{0}
\setcounter{table}{0}
\setcounter{page}{1}
\pagenumbering{arabic}
\noindent
\textbf{Supplemental Material}

\section{1. I\lowercase{nfrared microspectroscopy\\using a diamond anvil cell}}
A diamond anvil cell (DAC) with type IIa diamond anvils was used in the infrared microspectroscopy. Infrared spectra were collected on a Bruker Vertex 80v FT-IR spectrometer coupled to a Hyperion 2000 microscope, using a Globar source and an MCT detector. The measurement configuration is shown in Fig.~\ref{FIGS1}. The sample was pressed against the top diamond anvil and surrounded by a pressure-transmitting medium (KBr) elsewhere. The sample covered approximately half of the 300 $\mu$m diameter sample chamber in a stainless steel gasket. BiTeI is soft; when compressed in a DAC, it forms a shiny flat surface with the top diamond anvil to make specular reflections. For each pressure point, we first obtained the transmittance: the infrared beam was focused on the KBr next to the sample to collect a reference transmission spectrum $T_r$, and then on the sample to collect the sample transmission spectrum $T_s$; the ratio $T_s/T_r$ gave the sample transmittance. We next obtained the reflectance: the infrared beam was first focused on the sample to collect the sample reflection spectrum $R_s$, and then on the top diamond-air interface to collect a reference reflection spectrum $R_r$, yielding a ratio $R_1=R_s/R_r$. When all the pressure-dependent measurements were finished, we opened the DAC and took the top diamond anvil which had been in contact with the sample. After cleaning its culet, we measured the reflection spectrum from the top diamond-air interface ($R_r$) and from the culet-air interface through the anvil ($R_c$) to yield another ratio $R_2=R_c/R_r$. The sample reflectance at the sample-diamond interface was calculated as $R_1/R_2\cdot R_d =R_s/R_c\cdot R_d$, where $R_d$ is the known single-bounce reflectance of diamond at the diamond-air interface. Such a measurement scheme minimized possible systematic errors and the effects due to the absorption from diamond.

The spectra contain broad fringes due to Fabry-P\'{e}rot effects in the sample and fine fringes due to the same effects in the pressure-transmitting medium KBr. Because of our measurement scheme, both broad and fine fringes are present in the reflectance, but only fine fringes dominate in the transmittance. For reflectance, the sample spectrum $R_s$ contains Fabry-P\'{e}rot effects from both the sample and KBr, but the spectra $R_r$ and $R_c$ do not have fringes. The ratio $R_s/R_c$ therefore preserves both the broad and the fine fringes. Because the sample is absorbing, and because the beam has to traverse the sample twice to make the Fabry-P\'{e}rot effects in KBr observable in the measured reflectance, the corresponding fringe contrast is weak. Therefore these fine fringes are seen as small oscillations superimposed on a background of the much broader sample fringes. For transmittance, the reference spectrum $T_r$ was taken in the region next to the sample, where KBr is thicker than that underneath the sample by an amount equal to the sample thickness. This thickness mismatch gives rise to the dominant fine fringes in the transmittance.

\section{2. I\lowercase{nfrared data analysis}}
\begin{figure}[b]
\renewcommand{\thefigure}{S\arabic{figure}}
\includegraphics[scale=0.95]{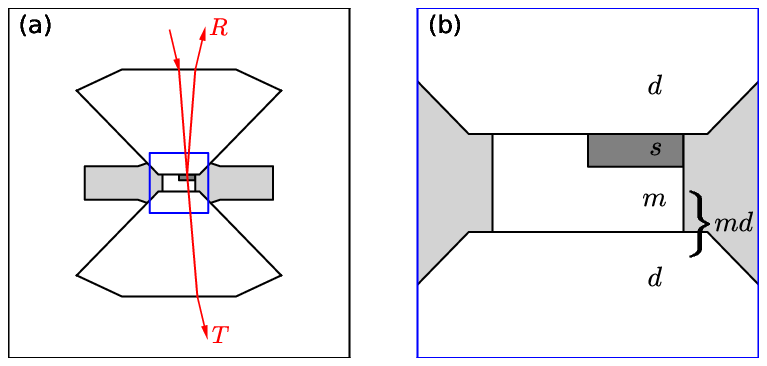}   
\caption{(a) Measurement configuration for infrared microspectroscopy in a diamond anvil cell. The arrowed lines indicate the beam paths for the reflected ($R$) and transmitted ($T$) signals. (b) Close-up of the part in the square in (a). $d$: diamond anvil. $s$: sample. $m$: pressure-transmitting medium. $md$: composite medium consisting of $m$ and $d$.} 
\label{FIGS1}
\end{figure}
The measured reflectance and transmittance can be analyzed by considering the sample (denoted as $s$) and the pressure-transmitting medium (denoted as $m$) as a double-layer slab, shown in Fig.~\ref{FIGS1}. This slab is bounded by the diamond anvils (denoted as $d$) on both sides. Since the fringes were removed from the spectra to simplify the analysis, one can add intensities rather than amplitudes from the multiple internal reflections in $s$ and $m$. Because the reflectance and transmittance of a single-layer slab is straightforward to obtain, we solve the double-layer slab problem following a two-step method \cite{Heavens1995}. (i) Solve the single-layer-slab problem, where the slab is $m$, bounded by $s$ and the bottom $d$. This gives the reflectance $R_{s \mhyphen md}$ and transmittance $T_{s\mhyphen md}=1-R_{s\mhyphen md}$ of a composite medium (denoted as $md$) consisting of $m$ and the bottom $d$, interfaced with $s$. (ii) Solve the single-layer-slab problem, where the slab is $s$, bounded by the top $d$ and the composite medium $md$. The reflectance and transmittance for the double-layer slab are
\begin{align}
R &= R_{sd}+\frac{(1-R_{sd})^2R_{s\mhyphen md}e^{-8\pi\kappa\nu x}}{1-R_{sd}R_{s\mhyphen md}e^{-8\pi\kappa\nu x}},\label{eqR}\tag{S1}\\
T &= \frac{(1-R_{sd})(1-R_{s\mhyphen md})e^{-4\pi\kappa\nu x}}{1-R_{sd}R_{s\mhyphen md}e^{-8\pi\kappa\nu x}},\label{eqT}\tag{S2}
\end{align}
in which
\begin{equation}
R_{s\mhyphen md} = R_{sm}+\frac{(1-R_{sm})^2R_{md}}{1-R_{sm}R_{md}}. \label{eqRsmd}\tag{S3}
\end{equation}
In Eqs.~\eqref{eqR},~\eqref{eqT}, and~\eqref{eqRsmd}, $R_{sd}=[(n-n_d)^2+\kappa^2]/[(n+n_d)^2+\kappa^2]$, $R_{sm}=[(n-n_m)^2+\kappa^2]/[(n+n_m)^2+\kappa^2]$, and $R_{md}=(n_m-n_d)^2/(n_m+n_d)^2$ are the single-bounce reflectance at the $s$-$d$, $s$-$m$, and $m$-$d$ interfaces, respectively. They involve the to-be-solved sample refractive index $n$ and extinction coefficient $\kappa$, and the known refractive index of diamond $n_d$ \cite{Edwards1981} and KBr $n_m$ \cite{Stephens1953}. $\nu$ is the wavenumber (defined as 1/wavelength with the unit cm$^{-1}$), and $x$ is the sample thickness. The determination of $x$ is discussed in the next section.

Using Eqs.~\eqref{eqR} and~\eqref{eqT}, we iteratively solve for the sample refractive index $n$ and extinction coefficient $\kappa$. These are used to calculate the optical conductivity. Its real part is plotted in Fig.~4 in the main text, showing the intraband and interband transitions reported previously in the 0.1--0.6~eV range (see Refs [29, 30] in the main text). The sudden drop of the low-energy conductivity above 2.20~GPa results in the fringe contrast improvement in reflectance and the increase in transmittance shown in Fig.~2 in the main text. The significant pressure-induced increase of $\sigma_1$ near 0.4~eV from interband transitions confirms the abrupt suppression of the transmittance at the same photon energies shown in Fig.~2 in the main text. The band gap closing and reopening associated with the TQPT is not discernible as interband features. This involves the complicated changes of the phase space and matrix element for the interband transitions under pressure. First-principles calculations taking into account the band-filling due to free carriers will be useful to clarify these interband features.

\section{3. S\lowercase{ample thickness}}
The thickness of the sample in the DAC can be determined from the fringes in the reflectance spectra. We Fourier transformed the raw reflectance spectrum at 0.17~GPa and filtered out the fringe signature due to KBr, leaving the broad fringes due to the sample. The broad fringes have a period of $\Delta\nu\approx 300$~cm$^{-1}$. Taking the refractive index of the sample calculated from the reflectance fit at this pressure in this photon energy range, $n\approx 4.1$, we determined the sample thickness $x=1/2n\Delta\nu\approx 4.1$~$\mu$m. Note that the sample refractive index $n$ is not a known quantity. We initially fitted the reflectance using the single-bounce reflectance formula for the diamond-sample interface, $R_{sd}=[(n-n_d)^2+\kappa^2]/[(n+n_d)^2+\kappa^2]$. The refractive index $n$ derived from the fit was used to calculate the sample thickness. The fit was then refined using Eq.~\eqref{eqR}. We checked the resulting $n$ and repeated these procedures, until a self-consistent solution was found. To determine the sample thickness at other pressures, we used the pressure dependence of the lattice parameter $c$ obtained from x-ray powder diffraction. 

To confirm the validity of the thickness estimation based on fringes, we applied this method to determine the thickness of the pressure-transmitting medium KBr. We obtained the fine-fringe period from the raw transmittance spectrum at 0.17~GPa, $\Delta\nu\approx83$~cm$^{-1}$. The refractive index of KBr is about 1.53 in the photon energy range where the fine fringes reside. Therefore the substrate thickness is roughly $1/(2\times 1.53\times 83$~cm$^{-1})=39$~$\mu$m. Note that we measured the thickness of the pre-indented part of the stainless-steel gasket to be $\sim$41~$\mu$m. This is reasonably close to the estimated KBr thickness.

\begin{figure*}[t!]
\renewcommand{\thefigure}{S\arabic{figure}}
\includegraphics[scale=0.9]{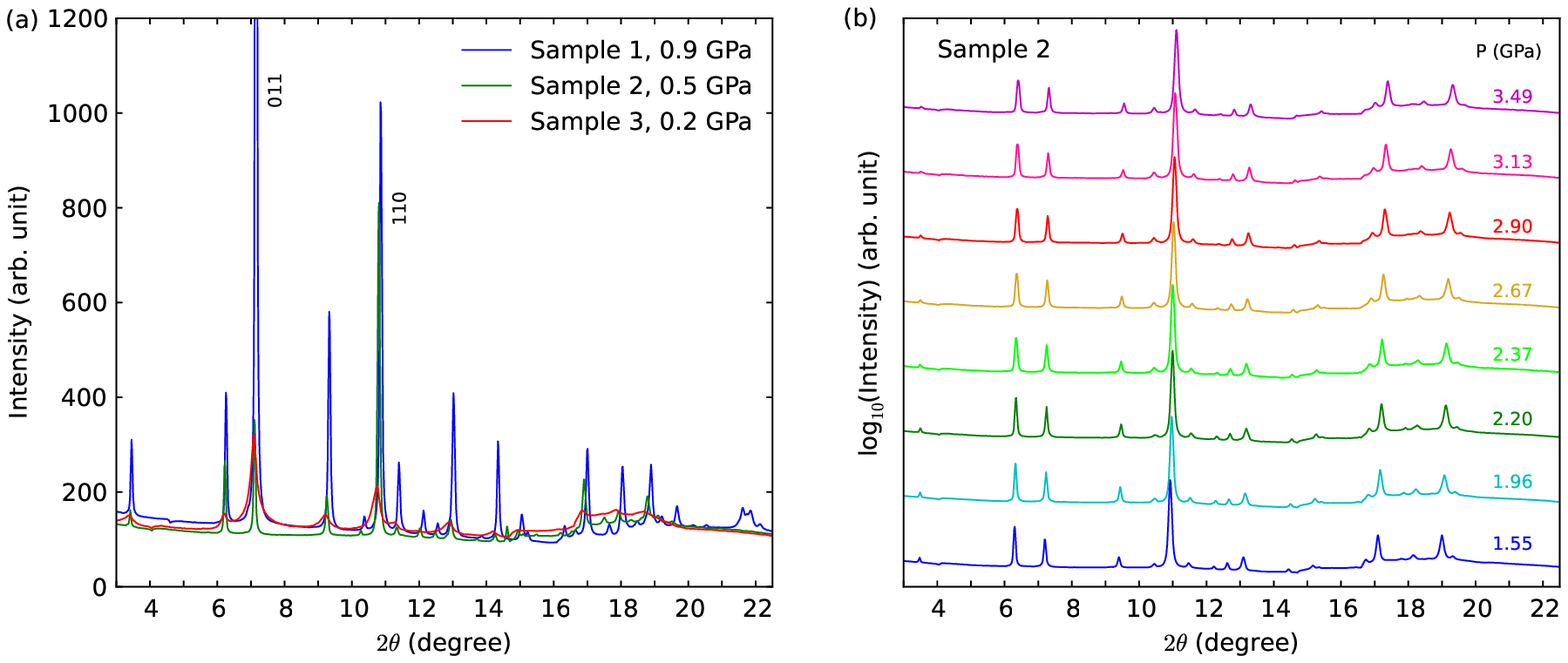}   
\caption{(a) X-ray powder diffraction patterns of BiTeI for three samples. The incident monochromatic x-ray wavelength is 0.4066~\AA~for Sample 1 and 0.4072~\AA~for Sample 2 and Sample 3. (b) X-ray powder diffraction patterns for Sample~2 from 1.55 to 3.49 GPa.} 
\label{FIGS2}
\end{figure*}

\begin{figure*}[t!]
\renewcommand{\thefigure}{S\arabic{figure}}
\includegraphics[scale=0.9]{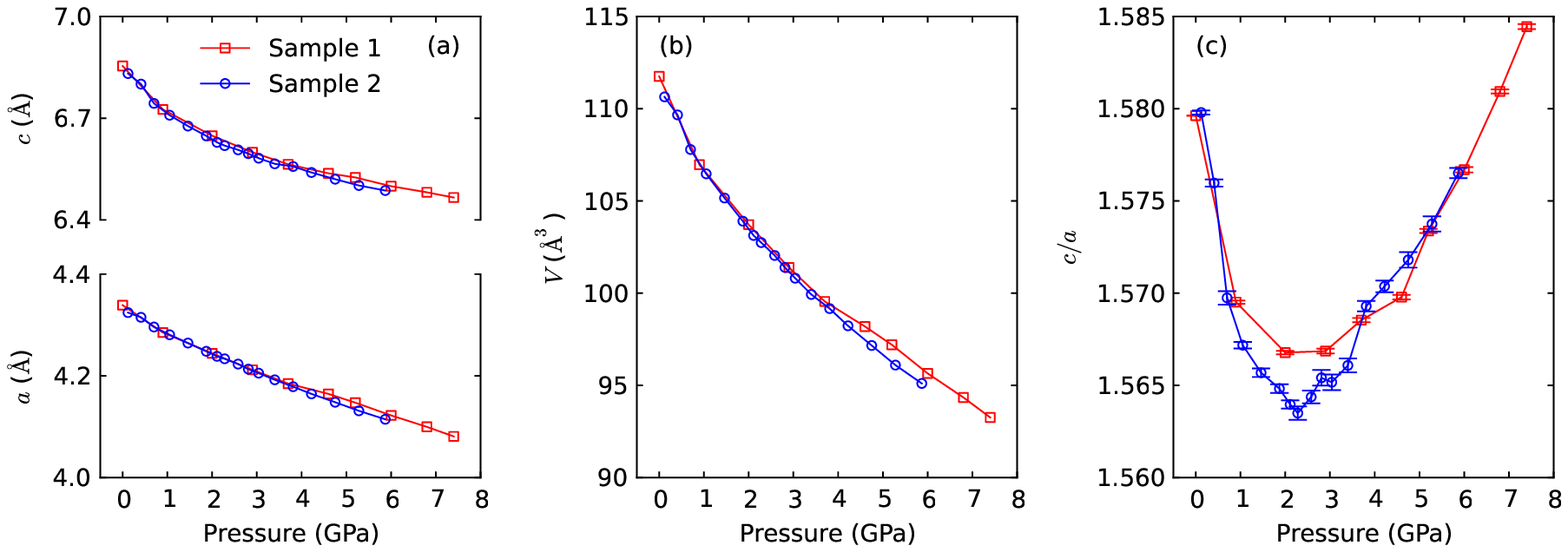}   
\caption{Pressure dependence of (a) the lattice parameters $c$ and $a$, (b) the unit-cell volume $V$, and (c) the ratio $c/a$ for Sample 1 (squares) and Sample 2 (circles). Error bars in (a) and (b) are within the symbol size. Error bars in (c) come from the uncertainties in $c$ and $a$.} 
\label{FIGS3}
\end{figure*}

\section{4. S\lowercase{upplementary} XRD \lowercase{data}}
To confirm the absence of any structural phase transition below 8~GPa, especially between 2--3~GPa where the topological quantum phase transition occurs, we performed additional x-ray powder diffraction measurements using smaller pressure steps. As shown in Fig.~\ref{FIGS2}(b), each Bragg peak systematically shifts to a larger $2\theta$ angle upon pressure increase. No additional or missing peaks were observed, confirming that BiTeI maintains its ambient-pressure structure between 2--3~GPa.

We note that sample preparations affect the powder diffraction data and the extracted lattice parameters. Due to its layered structure and its malleability, BiTeI cannot be easily ground into fine powder. The resulting XRD patterns show enhanced peak intensity along certain crystallographic directions [e.g. those labelled in Fig.~\ref{FIGS2}(a)], indicating preferred orientations. This factor is included in the Rietveld refinement of the data. Different levels of grinding was tested. The results show that sufficient grinding reduces the degree of preferred orientation but broadens the Bragg peaks. Such broadening progresses as the level of grinding is further increased, until the Bragg peaks merge to lose peak information [Sample 3 in Fig.~\ref{FIGS2}(a)]. For Rietveld refinement we chose the data sets for Sample 1 (presented in the main text) and Sample 2 (presented in this section) that do not show the peak broadening. 

Fig.~\ref{FIGS3} compares the analyzed results from Rietveld refinement for Sample 1 (taken from the main text) and Sample 2. The agreement is reasonable. The minimum in $c/a$ is reproduced. The quantitative difference could have many causes. For example, sample preparations may affect the data and hence the results, as discussed in the previous paragraph. Sample 1 and Sample 2 could be intrinsically different because they were taken from different parts of the same single crystal. The two measurements on Sample 1 and Sample 2 were taken five months apart, during which time the sample property could have slightly changed.

\end{document}